\documentclass[prl,9pt]{revtex4}

\usepackage{amsmath,amssymb}
\usepackage{verbatim}
\usepackage{graphicx}
\usepackage{hyperref}
\usepackage{color}
\usepackage{footnote}

\DeclareFontFamily{OT1}{rsfs}{}
\DeclareFontShape{OT1}{rsfs}{m}{n}{ <-7> rsfs5 <7-10> rsfs7 <10->rsfs10}{} 
\DeclareMathAlphabet{\mycal}{OT1}{rsfs}{m}{n}

\newcommand{\be}[1]{ \begin{equation}\label{#1} }
\newcommand{\ee}{\end{equation}}
\newcommand{\bea}[1]{\begin{eqnarray}\label{#1} }
\newcommand{\eea}{\end{eqnarray}}

\newcommand{\<}{\langle}
\renewcommand{\>}{\rangle}

\newcounter{rowcount}
\begin{document}

\title{One loop partition function of six dimensional conformal gravity using heat kernel on AdS}
\author{Iva Lovrekovi\'c}
\email{lovrekovic@hep.itp.tuwien.ac.at}
\affiliation{Institute for Theoretical Physics, Technische Universit\"at Wien, Wiedner Hauptstrasse 8--10/136, A-1040 Vienna, Austria}

\date{\today}

\preprint{TUW--13--xx}

\begin{abstract} 

We compute the heat kernel for the Laplacians of symmetric transverse traceless fields of arbitrary spin on the $AdS$ background in even number of dimensions using the  group theoretic approach introduced in \cite{Gopakumar:2011qs} and apply it on the partition function of six dimensional conformal gravity. 
The obtained partition function consists of the Einstein gravity, conformal ghost and two modes that contain mass.
\end{abstract}

\maketitle

\tableofcontents

\section{Introduction}

In order for the gravity theory to be renormalizable one has to include higher derivative terms \cite{Lu:2013hx} for which one can simultaneously demand to satisfy particular symmetry conditions such as conformal symmetry, that in four dimensions leads to conformal gravity.
Imposing the right boundary conditions to conformal gravity gives Einstein gravity  \cite{Maldacena:2011mk}, as well as using the procedure based on a tuning of a coefficient such that massive ghostlike spin-2 excitations are made massless \cite{Lu:2011ks}. 
 The latter was generalised to six dimensions, where the conformal gravity action consists of conformal anomalies considered in geometric description in arbitrary number of dimensions. With respect to the geometric description they belong to two different classes. First one that consists of Weyl invariants  that vanish in integer dimensions, usually referred to as type A, and the second one, type B, that is correlated to local conformal scalar polynomials of Weyl tensor and its derivatives. There is as well the third class that are trivial, local, anomalies, that can be removed using the local counter term  \cite{Deser:1993yx,Pang:2012rd,Erdmenger:1997gy,Bastianelli:2000rs}. 

Interest in six dimensional conformal gravity action originates in its numerous applications \cite{Deser:1993yx} and relevance for (0,2) theory \cite{Henningson:1998gx,Pang:2012rd} that can describe the connection of the conformal anomaly to tensionless strings \cite{Baulieu:2001pu}. General six dimensional theories have been studied in  \cite{Hull:2000zn,Hull:2000ih,Silverstein:1995re,Metsaev:2010kp}, classified according the geometry of its anomalies \cite{Deser:1993yx} and studied via analysis of its Lagrangian \cite{Metsaev:2010kp,Oliva:2010zd} and its contained divergencies \cite{Pang:2012rd}. 
The relation of the free-theory conformal anomaly to the Seeley-DeWitt coefficients \cite{Seeley:1967ea} has been discussed in \cite{Bastianelli:2000hi}, while the logarithmic divergence in one loop effective action on different backgrounds $S^6,CP^3,S^2\times S^4,S^2\times CP^2,S^3\times S^3$ and $S^2\times S^2\times S^2$ has been studied in \cite{Pang:2012rd}  \footnote{The heat kernel coefficients have been considered in \cite{Gilkey:1975iq}.}.

Here, we will consider the partition function of the conformal gravity in six dimensions for the combination of the conformal invariants such that their form coincides with the holographic anomaly of multiple coincident M5-branes \cite{Henningson:1998gx,Pang:2012rd}. 
Partition function plays as well an important role in the (2,0) conformal supergravity \cite{Beccaria:2015uta}.
Its investigation allows to derive the correction to thermodynamical quantities  studied in the framework of AdS/CFT correspondence \cite{Maldacena:1997zz}. In the original conjecture, the partition function of the d dimensional gravity theory in the bulk corresponds to the partition function of d-1 dimensional conformal field theory on the boundary of the AdS space when the holographic component is set to zero or infinity. 
 Large part of the research is in lower dimensions \cite{Maloney:2007ud,Bertin:2011jk,Gaberdiel:2010xv,Zojer:2012rj}. 
 For one loop partition function of 3D Einstein gravity, one obtains the result anticipated by Brown and Henneaux. The partition function consists of the sum over boundary excitations of $AdS_3$, Virasoro descendants of empty AdS space. The graviton one loop partition function of Einstein gravity  and Chern Simons gravity gives the product of Virasoro descendants from the Einstein gravity and an additional part, that has provided strong evidence the dual conformal field theory to TMG at the chiral point is logarithmic \cite{Gaberdiel:2010xv}. 

Our approach in the evaluation of the partition function is the group theoretic approach, that considers heat kernel on the background of AdS space. In addition to the explicit computation of the traced heat kernel for the Laplacians of the fields with arbitrary spin on odd dimensional AdS background in \cite{Gopakumar:2011qs}, we provide  the explicit expression for the heat kernel on $AdS_{2n}$ space and use that result to deduce the heat kernel on the $AdS_{6}$ and compute the partition function of the six dimensional conformal gravity.
We compare six dimensional conformal gravity partition function, with the three dimensional conformal gravity from the perspective of the modes it consists of \cite{Bertin:2011jk}. There, the one loop partition function consists of the one-loop partition function coming from Einstein gravity, conformal ghost and partially massless mode (which were in lower number of dimensions considered in \cite{Deser:2012qg}), and one massive mode \cite{Tseytlin:2013fca}.
The comparison in the sense of $AdS_3/CFT_2$ is special because two dimensional CFT allows for the computation of the partition function while analogous computation has not been done in higher dimensions. The higher dimensional CFTs, i.e. $CFT_5$, do not posses the same properties as $CFT_2$. 
For the generalisation of partition function computation from AdS and CFT side, to higher dimensions to be possible, it is essential to consider theory of the certain symmetry on the particular background, \cite{Beccaria:2014jxa}. 
 Partition function of the conformal spin s field has been considered from $CFT_d/AdS_{d+1}$ perspective, where $AdS_{d+1}$ has $S^1\times S^{d-1}$ boundary \cite{Beccaria:2014jxa}. In particular, the $d=4$ case exhibits property that partition function on $S^1\times S^3$ of the conformal higher spin (CHS) field is equal to partition function of the CHS  for the positive ground state energies 
on the $AdS$ background times two \cite{Beccaria:2016tqy}. That has not been observed in other dimensions.
 
 The heat kernel for the Laplacians of the fields  with even dimensional spins that are symmetric and transverse traceless can be used in computations for quantum anomalies \cite{Vassilevich:2003xt}, \cite{Gopakumar:2011qs}, particular  asymptotics of effective action and more \cite{Vassilevich:2003xt}. 
 In $AdS_{3}$ partition function of a higher spin theory was obtained as MacMahon function, which was done also in $AdS_5$ \cite{Gupta:2012he}. 

The first section introduces preliminaries and conventions after which we vary the action two times to obtain the linearised equations of motion and partition function in terms of  determinants. In section two, we  compute the heat kernel using the group theoretic approach on even dimensional $AdS$. We apply the general  expression from second section for computation of the partition function of the six dimensional conformal gravity, in section three. In section four we conclude.

\section{Conformal gravity in six dimensions}

 Conformal gravity in six dimensions can be written in terms of  invariants which are known as well as type B anomalies, \cite{Nutma:2014pua,Bastianelli:2000rs}, 
\begin{align}
I_1&=C_{\mu\nu\rho\sigma}C^{\mu\lambda\kappa\sigma}C_{\lambda}{}^{\nu\rho}{}_{\kappa} \\
I_2&=C_{\mu\nu\rho\sigma}C^{\rho\sigma\lambda\kappa}C_{\lambda\kappa}{}^{\mu\nu} \\
I_3&=C_{\mu\nu\rho\sigma}\left(\delta^{\mu}_\lambda\Box+4R^\mu_\lambda-\frac{6}{5}R\delta^\mu_\lambda\right)C^{\lambda\nu\rho\sigma}+\nabla_\mu J^\mu \label{invsb}
\end{align}
and the six dimensional Euler density, known as type A anomaly
\begin{equation}
E_6=\epsilon_{\mu_1\nu_1\mu_2\nu_2\mu_3\nu_3}\epsilon^{\rho_1\sigma_1\rho_2\sigma_2\rho_3\sigma_3}R^{\mu_1\nu_1}{}_{\rho_1\sigma_1}R^{\mu_2\nu_2}{}_{\rho_2\sigma_2}R^{\mu_3\nu_3}{}_{\rho_3\sigma_3},
\end{equation}
where tenor $J^a$ vanishes on Einstein spaces. %
It belongs to third class of the anomaly, trivial anomaly, and it has been computed and analysed in detail in \cite{Bastianelli:2000rs}. 
 One can obtain it as Weyl variation of the local functional which can be written in terms of total derivatives.  To cancel trivial anomalies, one has to add the local counterterms to the action. Term $\nabla_iJ^i$ can be written as
\begin{equation} 
\nabla_iJ^i=-\frac{2}{3}M_5-\frac{13}{3}M_{6}+2M_7+\frac{1}{3}M_8\label{tan}
\end{equation}
in the basis where we define 
 \begin{align}
 K_1&=R^3 && K_2=RR_{ab}^2 && K_3=RR_{abmn}^2 \nonumber \\
 K_4&=R_a^mR_m^iR_i^a && K_5=R_{ab}R_{mn}R^{mabn} && K_6=R_{ab}R^{amnl}R^b{}_{mnl}\nonumber \\
 K_7&=R_{ab}{}^{mn}R_{mn}{}^{ij}R_{ij}{}^{ab}&& K_8=R_{mnab}R^{mnij}R_i{}^{ab}{}_{j}&& K_{9}=R\nabla^2R \nonumber \\
 K_{10}&=R_{ab}\nabla^2R^{ab} && K_{11}=R_{abmn}\nabla^2R^{abmn} && K_{12}=R^{ab}\nabla_a\nabla_bR \nonumber \\
 K_{13}&=(\nabla_aR_{mn})^2 && K_{14}=\nabla_aR_{bm}\nabla^{b}R^{am} && K_{15}=(\nabla_iR_{abmn}^2) \nonumber \\
 K_{16}&=\nabla^2R^2 && K_{17}=\nabla^4R, \label{ks}
 \end{align}
and the $M_i$ for $i=5,6,7,8$ 
\begin{align}
M_5&=6K_6-3K_7+12K_8+K_10-7K_{11}-11K_{13}+12K_{14}-4K_{15}\\
M_6&=-\frac{1}{5}K_9+K_{10}+\frac{2}{5}K_{12}+K_{13}\\
M_7&=K_{4}+K_5-\frac{3}{20}K_9+\frac{4}{5}K_{12}+K_{14}\\
M_8&=-\frac{1}{5}K_9+K_{11}+\frac{2}{5}K_{12}+K_{15},\end{align}
and canceled by the local functionals, given in the Appendix. 
The general combination of the  invariants (\ref{invsb}) does not lead to the action whose variation gives equations of motion satisfied by Einstein metric. To use that kind of action we consider Lagrangian with a defined combination of invariants and the Euler term 
\begin{equation}
\mathcal{L}=\kappa\left( 4 I_1+I_2-\frac{1}{3}I_3-\frac{1}{24}E_6\right)
\end{equation}
whose variation leads to the equations of motion that are solved by the  Einstein metrics. That combination of the invariants and the Euler term has a form as holographic Weyl anomaly of number of coincident M5-branes \cite{Pang:2012rd} and leads to
\begin{equation}
\mathcal{S}=\kappa\int d^6x \sqrt{|g|} \left( 4 I_1+I_2-\frac{1}{3}I_3-\frac{1}{24}E_6\right).
\end{equation}
We are interested in the linearised equations of motion, so we vary the action and keep the bulk part of the result
\begin{equation} 
\delta^{(1)}S=\int d^6 x  \sqrt{|g|}EOM\delta g_{\mu\nu},\label{ac1}
\end{equation}
without keeping the boundary terms. The contribution to the linearised equations of motion will come only from the invariant $I_3$ of the action \cite{Tseytlin:2013fca,Beccaria:2015uta}. The second variation of the action (\ref{ac1}) leads to three terms, a term that contains the variation of the determinant multiplied by the equations of motion, the term that contains the second variation of the metric and it is set to zero, and the term that contains the variation of equations of motion.  Evaluated on shell this gives zero for the first term so that the total contribution that defines linearised equations of motion comes from  the third term. 
\begin{equation}
\delta^{(2)}S=\int d^6 x \sqrt{|g|} \delta EOM\delta g_{\mu\nu}.
\end{equation}

To express the linearised equations of motion, we denote first variation of the metric  with
\begin{align}\delta g_{\mu\nu}=h_{\mu\nu} & &\text{ and }& & \delta g^{\mu\nu}=-h^{\mu\nu}\label{def1var}\end{align}
where the metric $g_{\mu\nu}=\bar{g}_{\mu\nu}+h_{\mu\nu}$ is split in background $AdS$ metric $\bar{g}_{\mu\nu}$ and small perturbation $h_{\mu\nu}$ around the background, and indices are raised and lowered with  $\bar{g}$. 
In that notation, the second variation of the metric becomes
\begin{align}\delta^{(2)}g_{\mu\nu}=0, & & \delta^{(2)}g^{\mu\nu}=-\delta h^{\mu\nu}=2h^{\mu}{}_{\rho}h^{\rho\nu}.\end{align}
Since the tensors are evaluated on $AdS_6$ background, we can further simplify for respective spaces, and express the Riemann tensor in terms of the cosmological constant $\Lambda$ and the metric $\bar{g}_{ab}$, 
$R^{\mu\nu\rho\sigma}=\Lambda(-\bar{g}^{\mu \sigma}\bar{g}^{\nu\rho}+\bar{g}^{\mu\rho}\bar{g}^{\nu\sigma})$, that leads to  the Ricci tensor
$R^{\mu\nu}=5\Lambda \bar{g}^{\mu\nu}$ defined as well in terms of $\Lambda$ and $\bar{g}_{\mu\nu}$, and  the Ricci scalar determined by  
$R=30 \Lambda$. 
 In transverse traceless gauge of the metric, where $\nabla^{\mu}h_{\mu\nu}=0$ and $h_{\mu}^{\mu}=0$,  linearized equations of motion lead to the action 
\begin{equation} 
\delta^{(2)}S=\int d^6x\sqrt{-g}\left( 48 \Lambda^3 h^{\mu\nu}  -38 \Lambda^2 \nabla_{\rho}\nabla^{\rho}h^{\mu\nu} + 8 \Lambda \nabla_{\sigma}\nabla^{\sigma}\nabla_{\rho}\nabla^{\rho}h^{\mu\nu}  - \tfrac{1}{2} \nabla_{\lambda}\nabla^{\lambda}\nabla_{\rho}\nabla^{\rho}\nabla_{\sigma}\nabla^{\sigma}h^{\mu\nu} \right)\label{acti}.
\end{equation}
Here, $\nabla$ denotes covariant derivative with the respect to the background metric $\bar{g}$.

Our aim is to compute the one-loop partition function which we can write as a multiplication of three terms
\begin{equation}
Z_{1-\text{loop}}=\int \mathcal{D}\delta h_{\mu\nu}\times ghost \times exp(-\delta^{(2)}S),
\end{equation} 
the "ghost" term  that defines the determinants that we obtain when we eliminate the gauge degrees of freedom to which we conventionally refer to as ghost determinants, $\mathcal{D}{h_{\mu\nu}}$, which denotes the path integral over the perturbations $h_{\mu\nu}$ around the background thermal Euclidean $AdS_6$, and $exp(-\delta^{(2)}S)$, the exponential of the second variation of the action (\ref{acti}).

If we compute the second variation of the action using the decomposition of perturbations of the metric $h_{\mu\nu}$ 
\begin{equation}
h_{\mu\nu}=h^{\text{TT}}_{\mu\nu} + \tfrac{1}{6} \bar{g}_{\mu\nu} h +2\nabla_{(\mu}\xi_{\nu)},
\end{equation}
where $h^{TT}$ denotes transverse traceless part of the metric  $h_{\mu}^{\mu TT}=\nabla^{\mu}h_{\mu\nu}^{TT}=0$, h denotes a trace of the metric, and $\nabla_{(\mu}\xi_{\nu)}$ gauge parts, we obtain the same action as the above with $h_{\mu\nu}^{TT}$ on the place of the $h_{\mu\nu}$. 
The difference is that in the first case the conditions on the perturbation were imposed, while in the second case the metric was split but the trace has not been preliminary set to zero, and there was no condition on the transverse part of the perturbation. The fact that the final result is the same, is due to the gauge condition and conformal invariance of the metric.

The functional integral over gauge degrees of freedom, $\xi$ and h can be performed trivially. It gives an infinite constant which describes the volume of the gauge group, the differomorphism group, that has to be eliminated. To count the gauge-equivalent configurations only once, one divides the path integral measure by the gauge group volume. Convenient way to express that is in the terms of the determinant $Z_{gh}$ defined by the Jacobian that comes from the change of variables $h_{\mu\nu}\rightarrow(h^{TT}_{\mu\nu},h,\xi_{\mu})$
\begin{equation}
\mathcal{D}h_{\mu\nu}=Z_{gh}\mathcal{D}h_{\mu\nu}^{TT}\mathcal{D}\xi_{\mu}\mathcal{D}h, 
\end{equation}
with definition of the path integral measures for  tensors
$
1=\int D h_{\mu\nu} \text{Exp}\left( -\langle h,h\rangle \right),
$
vectors $
1=\int D \xi_{\mu} \text{Exp}\left( -\langle \xi,\xi\rangle \right) \label{defxi}
$
and scalars $
1=\int D s \text{Exp}\left( -\langle s,s\rangle \right) $
and $\langle...\rangle$, ultralocal invariant scalar products 
\begin{align}
\langle h,h'\rangle & =\int d^4 x \sqrt{g} h^{\mu\nu}h'_{\mu\nu} \\ \label{hh}
\langle \xi,\xi'\rangle & =\int d^4 x \sqrt{g} \xi^{\mu}\xi'_{\mu} \\
\langle s,s'\rangle & =\int d^4 x \sqrt{g} s s' .\\
\end{align}
Since one can decompose $\xi_{\mu} $ into \begin{equation}\xi_{\mu}(\xi^T,\sigma)=\xi^{T}_{\mu}+\nabla_{\mu}\sigma\end{equation} where $\nabla^{\mu}\xi_{\mu}^{T}=0$, he has to, in addition, take into account the Jacobian factor $J_1$ that corresponds to the change of the variables $\mathcal{D}\xi_{\mu}\rightarrow\mathcal{D}\xi_{\mu}^{T}\mathcal{D}s$ 
 \begin{align}
 1&=\int D\xi_{\mu}^TDs J_1\text{Exp}\left(-\int d^4x\sqrt{g}\xi_\nu(\xi^T,s)\xi^{\nu}(\xi^T,s) \right) \\
 &=\int D\xi_{\mu}^{T} Ds J_1\text{Exp} \left( -\int d^4x\sqrt{g}(\xi_{\nu}^{T}\xi^{T\nu}-s\nabla^2s) \right) \\
 &=J_1\left[ \det(-\nabla^2)_0 \right]^{-1/2},
 \end{align}
where index 0 denotes the scalar determinant. We will use indices 1 and 2 to denote the vector and tensor determinants, respectively. The decomposition of the metric now becomes
\begin{equation}
h_{\mu\nu}=h^{\text{TT}}_{\mu\nu} + \tfrac{1}{6} \bar{g}_{\nu\mu} h + \nabla_{\mu}\xi ^T_{\nu} + 2 \nabla_{\mu}\nabla_{\nu}s + \nabla_{\nu}\xi ^T_{\mu}.
\end{equation}
If we denote by $J_2$ the Jacobian factor obtained from the change of variables $h_{\mu\nu}\rightarrow(h^{TT},h,\xi^T,s)$ we find
\begin{equation}
1=\int J_2 \mathcal{D}h_{\mu\nu}^{TT}\mathcal{D}h\mathcal{D}\xi_{\mu}^T\mathcal{D}s \exp\left(-\left<h(h^{TT},h,\xi^T,s),h(h^{TT},h,\xi^T,s) \right>\right)
\end{equation} which reads\begin{align}
1 &=\int Dh_{\mu\nu}^{TT}D\xi_{\mu}^{T}DhDs J_2  \times  \\ \nonumber  & 
 Exp\left\{ -\int d^4x \sqrt{g} \left[ h^{\text{TT}}_{\mu\nu} h^{\text{TT}\mu\nu} + \tfrac{1}{6} h^2 -\xi ^{T \mu} (10 \Lambda  +2 \nabla_{\nu}\nabla^{\nu})\xi ^T_{\mu}  + s(20 \Lambda  \nabla_{\mu}\nabla^{\mu} + \tfrac{10}{3}  \nabla_{\nu}\nabla^{\nu}\nabla_{\mu}\nabla^{\mu})s \right] \right\},
\end{align}
and defines $J_2$
\begin{align}
\mathcal{D}h_{\mu\nu}&=J_2\mathcal{D}h_{\mu\nu}^{TT}\mathcal{D}h\mathcal{D}\xi_{\mu}^T\mathcal{D}s \\
J_2&=\left[ \det(\nabla^2)_0 \right]^{1/2}[\det(5\lambda+\nabla^2)_1]^{1/2}[\det(6\lambda+\nabla^2)_0]^{1/2} \label{j6}
\end{align}
 where we have used the property that  $\det A \cdot \det B=\det AB$. However in the infinite dimensional case that does not have to be possible if the determinants are defined independently through for e.g. zeta functions. In quantum gravity this is usually treated such that determinants of higher order operators are {\it defined} by reducing them to the products of lower order operators \cite{Gaberdiel:2010xv},\cite{Elizalde:1997nd}. To obtain the ghost determinant we have to divide the above equation (\ref{j6}) with the Jacobian that corresponds to the change of variables $D\xi_{\mu}\rightarrow D\xi_{\mu}^{T}Ds$
\begin{equation}
Z_{gh}=\frac{J_2}{J_1}=[\det(5\lambda+\nabla^2)_1]^{1/2}[\det(6\lambda+\nabla^2)_0]^{1/2}. \label{zgh}
\end{equation}
We can now write the one loop partition function of conformal gravity action in six dimensions 
\begin{equation}
Z_{1-\text{loop}}^{(6)}=Z_{gh}\int D h_{\mu\nu}^{TT}Exp(-\delta^{(2)}S)=\frac{[\det(-\nabla^2-5\lambda)_1]^{1/2}[\det(-\nabla^2-6\lambda)_0]^{1/2}}{[\det(-\nabla^2+2\lambda)_2]^{1/2}[\det(-\nabla^{2}+6\lambda)_2]^{1/2}[\det(-\nabla^2+8\lambda)_2]^{1/2}}\label{zcgdet}
\end{equation}
and identify the one loop partition function of Einstein gravity as 
 \begin{equation}
 Z_{EG,1-\text{loop}}^{(6)}=\frac{[\det(-\nabla^2-5\lambda)_1]^{1/2}}{[\det(-\nabla^2+2\lambda)_2]^{1/2}}
 \end{equation}
the expression that agrees with the partition functions from \cite{Gupta:2012he,Giombi:2014yra}. 
The expression (\ref{zcgdet}) therefore consists of Einstein gravity $ Z_{EG}^{(6)}$, the scalar determinant in the numerator that denotes the contribution from the conformal ghost $[\det(-\nabla^2-6\lambda)_0]^{1/2}$, and two additional determinants in the denominator that correspond to modes which have mass $[\det(-\nabla^{2}+6\lambda)_2]^{1/2}[\det(-\nabla^2+8\lambda)_2]^{1/2}$. The first determinant denotes partially massless mode while the second one belongs to the massive mode.  It has been studied in a paper that investigates the quantum conformal higher spin (CHS) models where one obtains the partition function written in terms of determinants, that agrees with (\ref{zcgdet}) in the six dimensions for spin two case \cite{Tseytlin:2013fca}. 

We evaluate the above partition function of conformal gravity in six dimensions computing the heat kernel using the group theoretic approach.

\section{Heat Kernel for Partially Massless STT Field}

The determinants that appear in the partition function (\ref{zcgdet}) contain only symmetric transverse traceless fields (STT). %
To compute the determinant of STT field we have to evaluate the corresponding heat kernel. We could solve it on a manifold $\mathcal{M}$ via the appropriate heat equation by direct evaluation and construction of the eigenvalues and eigenfunctions for the Laplacian of the spin-S, and via computation of the resulting sum. However, for $\mathcal{M}$ a homogeneous space, 
 more convenient computation is using group theoretic techniques \cite{Gopakumar:2011qs} that use evaluation of the heat kernel on Euclidean $AdS_{2n+1}$ background, i.e. odd-dimensional hyperboloids, which can be analytically continuated to even-dimensional ones.  One has to keep in mind that the case of even-dimensional hyperboloids is more subtle and take one of the two following approaches. Evaluate the heat kernel using group theoretic techniques on even-dimensional $AdS$ space directly, the approach we follow here, or preform the analytic continuation from the odd-dimensional $AdS$.

Let us first remember the basic equations that connect the heat kernel and partition function. The trace of a heat kernel is related to the one-loop partition function via \begin{equation}
ln Z^{(S)}_{1-\text{loop}}=ln\det(-\Delta_{(S)})=Tr ln(-\Delta_{(S)})=-\int\limits_0^{\infty}\frac{dt}{t}Tre^{t\Delta_{(S)}}
\end{equation}
with the trace of the heat kernel defined as $Tr e^{t\Delta_{S}}$ under the integral, the Laplacian that defines a spin-S field on a manifold $\mathcal{M}$ denoted with $\Delta_{(S)}$, and the integration over the proper time $t$.
The heat kernel between two points x and y is defined using the normalised eigenfunctions $\psi_{n,a}^{(S)}$ that belong to $\Delta_{(S)}$, and the spectrum of eigenvalues $E_n{}^{(S)}$
\begin{equation}
K_{ab}{}^{(S)}(t;x,y)=\left\< y,b |e^{t\Delta_{(S)}}|x,a \right\>=\sum\limits_{n}\psi_{n,a}^{(S)}(x)\psi_{n,b}^{(S)}(y)^*e^{tE_n^{(S)}}\label{hkgen}
\end{equation}
where indices $a$ and $b$ are local Lorentz indices of the field. 
Taking the trace over the spin and space-time indices one defines the trace of a heat kernel 
\begin{equation}
K^{(S)}(t)\equiv Tr e^{t\Delta_{(S)}}=\int_{\mathcal{M}}d^{d+1}x\sqrt{g}\sum\limits_aK_{aa}^{(S)}(t;x,x) \label{trhkgen}.
\end{equation}

To evaluate the heat kernel on symmetric space such as sphere or hyperbolic space (which is equal to Euclidean AdS) using the group theoretic approach, one has to preform the harmonic analysis on homogeneous spaces (for example $S^{2n}$),  quotients of that spaces, and analytically continuate 
 to Euclidean hyperboloids. 
Analytic continuation from sphere for odd-dimensional hyperboloids contains only "principal series" contribution to the heat kernel. 
For even-dimensional hyperboloids there can in general exist contribution from the discrete series as well, however it does not contribute for the STT fields. 
The exception appears for two dimensional hyperboloids. In $AdS_2$ there is an additional contribution from the discrete series, even in the case of STT tensors \cite{Gopakumar:2011qs,Camporesi:1994ga}. 

 Evaluation of the partition function, as we will see, will lead to the result that contains the log divergence which is automatically standardly renormalised as described in \cite{Vassilevich:2003xt}. I.e. the divergent term is subtracted according to standard renormalization procedure. Stating that, we do not consider that divergence in our computations. 
Analogously that contribution therefore does not appear in any of the  terms added or subtracted from the action.

\subsection{Traced heat kernel for even-dimensional hyperboloids}

To compute the heat kernel between the two points x and y we have to know the eigenfunctions ($\psi_{n,a}^{(S)}(x)$)  of the Laplacian $\Delta_{(S)}$ with energy spectrum $E_n^{(S)}$ for a spin field S defined on manifold $\mathcal{M}$  that is in our approach, initially a 2n sphere $S^{2n}\simeq SO(2n+1)/SO(2n)$. From which one induces eigenfunctions and eigenvalues on hyperbolic $\mathbb{H}_{2n}$, essentially Euclidean $AdS_{2n}$.

In the group theoretic language that means that we have two compact Lie groups G and H such that $H\in G$. 
We define representation $R$ of $G$ and its vector space $V_R$ of a dimension $d_R$ and a unitary irreducible representation $S$ of $H$ with vector space $V_S$ and dimension $d_S$. The indices on $V_S$ subspace of $V_R$ we denote with $a$ and the ones on $V_R$ with $I$.  The form G/H is coset space defined  by $G/H=\{gH\}$ for $g\in G$ which is called quotienting with the right action of H on G (the quotienting with left action we denote with $\Gamma\setminus G/H$) and the section 
 that determines  the eigenfunctions  is defined  via the map $\sigma:G/H\rightarrow G$ for which there is defined projection map $\pi:G\rightarrow G/H$ such that $\pi\circ\sigma= e$ for $e$ identity in G.
 We will use the sections $\sigma(gH)=g_0$ for $g_0$ an element of the coset gH chosen to obey rules mentioned in \cite{Gopakumar:2011qs}.
 
 The eigenfunctions of spin S Laplacian on the sphere are then defined via matrix elements 
 \begin{equation}
 \psi_a^{(S)I}(x)=\mathcal{U}^{(R)}(\sigma(x)^{-1})_{a}^I. \label{ef1}
 \end{equation}
If we denote the index $n$ in (\ref{trhkgen}) and (\ref{hkgen}) with labels $(R,I)$, using the eigenfunction (\ref{ef1}) we can write (\ref{hkgen}) as

\begin{align}
K_{ab}(x,y;t)=\sum\limits_{R}a_R^{(S)}\mathcal{U}^{(R)}(\sigma (x)^{-1}\sigma(y))_a{}^{b}e^{tE_R^{(S)}}\label{kab}
\end{align}
where $a_R^{(S)}=\frac{d_R}{d_S}\frac{1}{V_{G/H}}$, for $V_{G/H}$ volume of the $G/H$ space. In (\ref{kab}), we do not write index I for the energy eigenvalues because we are interested in the coset spaces SO(N+1)/SO(N) and SO(N,1)/SO(N) for which representation R contains S only once since the eigenfunctions with same R but different I are degenerate \cite{Gopakumar:2011qs}.

 (\ref{trhkgen}) is now 
\begin{equation}
K^{(S)}(x,y;t)\equiv\sum\limits_{a=1}^{d_S}K_{aa}^{(S)}(x,y;t)=\sum\limits_Ra_R^{(S)}\sum\limits_{a=1}^{d_S}\langle a,S| \mathcal{U}^{(R)}(\sigma (x)^{-1}\sigma (y)) | a,S\rangle e^{tE_R^{(S)}} \label{css}
\end{equation}
where the second summation defines the trace of $\mathcal{U}$ over indices $a$ and we can write
\begin{equation}
K^{(S)}(x,y;t)=\sum_Ra_R^{(S)}Tr_S\mathcal{U}(\sigma(x)^{-1}\sigma(y))e^{tE_R^{(S)}}
\end{equation}
the traced heat kernel over G/H.
To determine the traced heat kernel on the thermal quotient of $S^{2n}$ we 
have to consider the quotient $\Gamma\backslash G/H$ for discrete group $\Gamma$ isomorphic to $\mathbb{Z}_N$ for the thermal quotient of the N-sphere, that can be embedded into $G$.  The choice of section that is {\it compatible} with quotienting by $\Gamma$ is defined so that if $\gamma \in \Gamma$ acts on $x=gH\in G/H$ via $\gamma:gH\rightarrow\gamma\cdot gH$, and a section $\sigma(x)$ is compatible with the quotienting $\Gamma$ if and only if 
\begin{equation}\sigma(\gamma(x))=\gamma\cdot\sigma(x)\label{compat}.\end{equation}

This allows us to use method of images \cite{David:2009xg} and the relation

\begin{equation}
K_{\Gamma}^{(S)}(x,y;t)=\sum\limits_{\gamma\in\Gamma}K^{(S)}(x,\gamma(y);t)\label{mirror}
\end{equation}
to compute the heat kernel $K_{\Gamma}^{(S)}$ between two points x and y on $\Gamma\backslash G/H$. In another words, if we fix, for example,  the point x and sum over the images of the point y, we can write
\begin{equation}
K_{\Gamma}^{(S)}(t)=\sum_{m\in \mathbb{Z}_N}\int_{\Gamma\backslash G/H}d\mu(x)\sum_{a}K_{aa}(x,\gamma^m(x);t)\label{mirim}
\end{equation}
where we have taken into account that  $\Gamma\simeq\mathbb{Z}_N$. Using the properties of integral over the quotient space, (\ref{mirim}) can be written as \cite{Gopakumar:2011qs}
\begin{equation}
K_{\Gamma}^{(S)}=\frac{\alpha_1}{2\pi}\sum\limits_{k\in\mathbb{Z}_N}\sum\limits_{R}\chi_{R}(\gamma^k)e^{t E_R(S)} \label{thk}
\end{equation}
with $\frac{\alpha_1}{2\pi}$ a volume factor for the thermal quotient $\gamma$. $\chi_R$  is the character in the representation R, $E_R(S)$ eigenvalue of the spin-S Laplacian on $S^{2n}$ and
the quotient $\gamma$, is the exponential of the Cartan generators of R. The representations R that are included in SO(2n+1) are representations that contain $S$ when restricted to the $SO(2n)$. 

To evaluate this, one has to know the characters of SO(2n+1) and eigenvalues $E_R$ \cite{Camporesi:1994ga}
\begin{equation}
E_{R,AdS_{2n}}^{(S)}=-(\lambda^2+\rho^2+s)
\end{equation}
where $\rho\equiv\frac{N-1}{2}$, for $N$ dimension of space. 

\subsection{Heat kernel on the even dimensional AdS}

The above analysis describes the computation of the heat kernel on a  compact symmetric space ($S^{2n}$) and the left quotients of that space.
The characters that we have to evaluate (for SO(2n+1)) are for the  compact symmetric space, and we want to extend  the analysis on the hyperbolic space $\mathbb{H}_{N}$ in order to consider the heat kernel on Euclidean $AdS$. Euclidean $AdS$ is the hyperbolic space $\mathbb{H}_N$, 
\begin{equation}
\mathbb{H}_N\approx SO(N,1)/SO(N)
\end{equation}
 where  N is the dimension of the space and one computes the heat kernel on $\mathbb{H}_N$ by writing a section in $SO(N,1)$ obtained via analytical continuation from $SO(N+1)$. In terms of the coordinates and explicit line element  we consider the coordinates of $S^{2n}$ and the metric 
\begin{equation}
ds^2=d\theta^2+\cos^2\theta d\phi_1^2+\sin^2\theta d\Omega_{2n-2}^2
\end{equation}
which after analytic continuation 
\begin{align}
\theta\rightarrow-i\rho, && \phi_1\rightarrow i t,\label{acn}
\end{align}
with $\rho$ and $t$ that contain values in $\mathbb{R}$,
becomes
\begin{equation}
ds^2=-(d\rho^2+\cosh^2\rho dt^2+\sinh^2\rho d\omega_{2n-2}^2).
\end{equation}.

In terms of the Lie algebra we continue SO(N+1) to SO(N,1) by choosing one axis to be time direction e.g., axis "1" and preforming the continuation of the generators $Q_{1j}\rightarrow iQ_{1j}$ of the corresponding algebras. 

The expressions for eignevalues and eigenfunctions are equal as in the compact case, where they are expressed in terms of matrix elements of unitary representations  of G.
However, now the unitary representations G that  define the matrix elements are  infinite dimensional because G is non-compact. These representations have been classified for SO(N,1).

We need unitary representations of SO(N,1) that have unitary representations of $SO(N)$. For even dimensional hyperboloids, with $N=2n$ these unitary representations are representations of principal series of $SO(2n,1)$
\begin{align}
R=(i\lambda,m_2,m_3,...,m_n), && \lambda\in\mathbb{R}, && m_2\geq m_3\geq...\geq m_n
\end{align}
for $m_i$ non-negative (half-)integers, where we  usually denote $m_2,m_3,...,m_n$ by $\vec{m}$. The principal series representations contain S of $SO(2n)$ according to branching rules \cite{Gopakumar:2011qs}
\begin{equation}
s_1\geq m_2\geq s_2\geq ...\geq m_n \geq |s_n|.
\end{equation}
These branching rules further simplify for $STT$ tensors because $m_2=s$, while $m_i=s_{i-1}=0$ for $i>2$ \footnote{There is an exception for n=1 when $|m_2|=s$}. The highest weight of the representation is in that case (s,0,...,0).

We are actually interested in computing the traced heat kernel of a tensor on the thermal quotient $AdS_{2n}$, that is a hyperbolic space $\mathbb{H}_{2n}$ with $\mathbb{Z}$ identification of coordinates
\begin{align}
t \sim t+\beta, & &\beta=i\alpha_1
\end{align}
where $\beta$ is inverse temperature, which corresponds to analytic continuation by (\ref{acn}) of phases of thermal quotient $\gamma$ on the sphere.  As in prescription \cite{Gopakumar:2011qs} we want to continue the heat kernel on the thermal sphere to thermal $AdS$ so that $\Gamma\approx\mathbb{Z}$ while for the sphere it was $\mathbb{Z}_N$. That means that on the place of the  character of SO(2n+1) in (\ref{thk}) we have the Harish-Chandra, i.e. global character of the group SO(2n,1)  (non-compact).
The traced heat kernel on thermal $AdS_{2n}$ becomes 
\begin{equation} 
K^{(S)}(\gamma,t)=\frac{\beta}{2\pi}\sum\limits_{k\in\mathbb{Z}}\sum\limits_{\vec{m}}\int\limits_{0}^{\infty}d\lambda\chi_{\lambda,\vec{m}}(\gamma^k)e^{tE_R^{(S)}}\label{thkads}
\end{equation}
while the global characters of the generators of the SO(2n,1) group are known from  \cite{hirai}. 
Reading out the character leads to
\begin{equation}
\chi(\beta,\phi_1,\phi_2,...,\phi_n)=\frac{\cos(\beta\lambda)\chi^{SO(2l+1)}(\gamma)}{2^{2l}\sinh^{2l+1}\left(\frac{\beta}{2}\right)}
\end{equation}
where  we have taken into consideration that for the thermal quotient $\beta\neq0$ and $\phi_i=0$  $ \forall i $ and $l=n-1$ \cite{hirai}. 
Prescription is to insert this into equation for the traced heat kernel on $AdS_{2n}$ (\ref{thkads}) for $\vec{m}=(m_2,...,m_n)$, highest weights of $\chi^{SO(2l+1)}_{\vec{m}}$ and integrate. However, for the STT fields the highest weights become $\vec{m}=(s,0,..,0)$, which we denote with $(s,0)$, and equation (\ref{thkads}) simplifies to
\begin{equation}
K^{(S)}(\beta,t)=\frac{\beta}{2^{2l+1}\sqrt{\pi t}}\sum\limits_{k\in\mathbb{Z}_+}\chi^{SO(2l+1)}_{(s,0)}\frac{1}{\sinh^{2l+1}\frac{k\beta}{2}}e^{-\frac{k^2\beta^2}{4t}-t(\rho^2+s)}.
\end{equation}
Where the summation is over $\mathbb{Z}_+$ since we do not include divergent $k=0$ term that appears because the volume of $AdS$ is infinite and we can integrate the coincident heat kernel  on the full $AdS_{2n}$.  That term does not depend on $\beta$ so we can absorb it into parameters of the theory. 


\section{The one-loop partition function of conformal gravity in six dimensions}

For the evaluation of the traced heat kernel we will need to evaluate the integral

\begin{equation}
\int \frac{dt}{t^{3/2}}e^{-\frac{a^2}{4t}-b^2 t}=\frac{2\sqrt{\pi}}{a}e^{-ab}
\end{equation}
which enters in the calculation of the one-loop determinant via
\begin{equation}
-\log\det(-\Delta_{(S)}+m_S^2)=\int\limits_0^{\infty}\frac{dt}{t}K^{(S)}(\beta,t)e^{-m_S^2t}.
\end{equation}
 One then obtains for the  STT tensors 
 
 \begin{equation}
-\log\det(-\Delta_{(S)}+m_S^2)=\frac{1}{2^{2l}}\sum\limits_{k\in\mathbb{Z}_+}\chi^{SO(2l+1)}_{(s,0)}\frac{1}{\sinh^{2l+1}\frac{k\beta}{2}}\frac{1}{k}e^{-k\beta\sqrt{\rho^2+s+m_S^2}},
\end{equation}
which we can rewrite into the heat kernel for the symmetric transverse traceless tenors on the even dimensional spaces
 \begin{equation}
-\log\det(-\Delta_{(S)}+m_S^2)=\sum\limits_{k\in\mathbb{Z}_+}\chi^{SO(2l+1)}_{(s,0)}\frac{2}{(1-e^{-k\beta})^{2l+1}e^{k\beta l }e^{\frac{k\beta}{2}}}\frac{1}{k}e^{-k\beta \sqrt{\rho^2+s+m_S^2}}.\label{main1}
\end{equation}
From (\ref{main1}) and (\ref{zcgdet}), we can read out the partition function of conformal gravity in six dimensions

\begin{align}
\log Z_{1-\text{loop}}^{(6)} =&\sum_{k\in\mathcal{Z_+}}
\frac{-e^{-\frac{5}{2}k\beta}}{k(1-e^{-k\beta})^5}(\chi_{(1,0)}^{SO(5)} e^{-\frac{7}{2}k\beta}+\chi_{(0,0)}^{SO(5)} e^{-\frac{7}{2}k\beta} \\
 &-\chi_{(2,0)}^{SO(5)} e^{-\frac{5}{2}k\beta}-\chi_{(2,0)}^{SO(5)} e^{-\frac{3}{2}k\beta}-\chi_{(2,0)}^{SO(5)} e^{-\frac{1}{2}k\beta} ). \label{z61}
\end{align}
To evaluate (\ref{z61}) we have to determine the characters of the SO(5)  group which have been listed in \cite{Tseytlin:2013jya}, and they are given in dependency of the spin of the field under consideration $\chi_{(s,0)}^{SO(5)}=\frac{1}{6}(2s+3)(s+2)(s+1)$, such that (\ref{z61}) becomes
\begin{align}
\log Z_{1-\text{loop}}^{(6)}=\sum_{k\in\mathbb{Z}_{+}} \frac{-2 q^{3k}}{k(1-q^k)^5}\left(3 q^{3k}-7q^{2k}-7q-7\right)
\end{align}
where we have denoted $q=e^{-\beta}$.
The terms in the partition function coming purely from Einstein gravity in six dimensions are \begin{equation}
\log Z_{EG,1-\text{loop}}^{(6)}=\sum_{k\in\mathbb{Z}_{+}} \frac{- q^{5k}}{k(1-q^k)^5}\left(5 q^k-14\right)\label{zeg},
\end{equation}
 while the remaining part comes from the conformal ghost, partially massless mode and massive mode
 \begin{align}
\log Z_{\text{diff},1-\text{loop}}^{(6)}=\sum_{k\in\mathbb{Z}_{+}} \frac{- q^{3k}}{k(1-q^k)^5}\left(q^{3k}-14q^k-14\right)
\end{align}
where the first term in the brackets originates from the conformal ghost. 
   The partition function for the free conformal higher spin theory has been considered in \cite{Tseytlin:2013jya} for the odd number of general dimensions and for the lower dimensions on $S^1\times S^{d-1}$ backgrounds which allows comparisons with the partition function of conformally invariant theories in these dimensions, keeping in mind the background metrics.
    The partition function of Einstein gravity has been considered in the terms of higher spin theories in \cite{Giombi:2014yra} in which one can find the agreement (\ref{zeg}) for the spin two case and the five dimensional conformal field theory. In three dimensions, the one loop partition function has been evaluated for the conformally invariant theory \cite{Bertin:2011jk}. If we compare the obtained one loop partition function to ours, we notice analogous contributions from Einstein gravity, partially massless response, and the conformal ghost. 

\section{Discussion}

In the article we compute and provide the explicit expression for the heat kernel for the Laplacians of symmetric transverse traceless tensors of arbitrary spin on the even dimensional Anti-de Sitter backgrounds. Group theoretic approach that we use is 
the analytic continuation from the spherical to hyperbolic background 
on even dimensional $AdS$ spaces,
on which we compute the traced heat kernel 
using the Harish-Chandra character of the principal series of SO(2n,1). The final expression of the one loop determinant of the spin s particle for STT tensors on the even dimensional backgrounds,  compares to the analogous one loop determinant on the odd dimensional spaces. Setting $l$ from the expression for the one loop determinant for the STT tenors on odd spaces to the $l+\frac{1}{2}$, leads to the expression for the one loop determinant on even spaces. 

In the computation of the partition function for six dimensional conformal gravity we obtain the expression in terms of one loop determinants that belong to symmetric transverse traceless fields. This allows us to use the expression for the heat kernel for the Laplacians of STT tensors  evaluated on the even dimensional spaces to compute the partition function consisted of the scalar, vector and tensor part with the non-vanishing mass. This structure is already visible from the partition function written in terms of the determinants, from which we can clearly recognise the Einstein gravity partition function, partition function that originates from the conformal ghost, and the partition function that appears due to partially massless and massive mode. This structure is as well visible in the conformal theories in lower dimensions. In \cite{Bertin:2011jk}, conformal gravity in three dimensions  shows the structure consisted of the Einstein gravity partition function, the one that originates from conformal ghost and the one due to partially massless mode, however in our case there is in addition one massive mode present. 
The result can be as well compared with the partition functions computed via operator counting method for the lower dimensions and higher odd dimensions on the $S^1\times S^{d-1}$ background \cite{Tseytlin:2013jya}.

Knowing the partition function of the theory one obtains the information about the canonical ensemble entropy, and can compute micro canonical entropy which counts the number of states N(M,J) at fixed energy M and angular momentum J \cite{Maloney:2007ud}.
The unique linear combination of invariants that we have studied allows for Einstein metrics to also be solutions of theory. The closed form class of solutions that corresponds to Schwarzshild-AdS metrics and their spherically symmetric conformal rescaling shows first law for five-parameter family of solutions \cite{Lu:2013hx}. Knowing the partition function one can further study the solutions by computing the correction to thermodynamical quantities.

The result for the heat kernel for the Laplacian of the STT tensor of arbitrary spin on even AdS background can be used in the computations that require evaluation of respective tensor fields. These are in studying one-loop divergencies, anomalies and given asymptotics of the effective action \cite{Vassilevich:2003xt}. Since the quantum level of the string theory sigma model whose target space is AdS is not that well known, one can also get an insight into the spectrum of the gauge theory and worldsheet structure \cite{Gopakumar:2011qs,polchinski1986}. Analogously to analysis of the partition function on odd dimensional $AdS$ spacetimes in \cite{Tseytlin:2013jya} one can consider the Laplacians of the STT operators obtained from the theory that is gauge and Weyl invariant \cite{Joung:2012qy} on even dimensional spaces and study higher spin theories.  
In this case one would have to recognise that partition function is consisted of those that would appear from Vasiliev theory \cite{Bekaert:2005vh}.

Computation of the partition function allows the further possible studies via detailed analysis in the AdS/CFT framework, and in computation of the thermodynamical quantities.

 \section{Acknowledgements}
 
 I would like to thank Daniel Grumiller, Dmitri Vassilevich for the discussions and Arkady Tseytlin for communications. The work was funded by the {\it Forschungsstipendien 2015}  of {\it Technische Universit\"at Wien}.
 
 \section{Appendix A: Local functionals that generate trivial anomalies}
 The local functionals that generate trivial anomalies are defined with
 \begin{align}
 \mathcal{M}_i&=\int d^6x\sqrt{g}\sigma(x)M_i\\
 \mathcal{K}_i&=\int d^6x\sqrt{g}K_i
 \end{align}
in which, in notation of \cite{Bastianelli:2000rs}
\begin{align}
\mathcal{M}_5&=\delta_{\sigma}\bigg( \frac{1}{30}\mathcal{K}_1-\frac{1}{4}\mathcal{K}_2+\mathcal{K}_{6}\bigg), \mathcal{M}_6=\delta_{\sigma}\bigg( \frac{1}{100}\mathcal{K}_1-\frac{1}{20}\mathcal{K}_2\bigg),  \nonumber\\
\mathcal{M}_7&=\delta_{\sigma}\bigg(\frac{37}{6000}\mathcal{K}_1-\frac{7}{150}\mathcal{K}_2+\frac{1}{75}\mathcal{K}_3-\frac{1}{10}\mathcal{K}_5-\frac{1}{15}\mathcal{K}_6\bigg), \mathcal{M}_8=\delta_{\sigma}\bigg(\frac{1}{150}\mathcal{K}_1\-\frac{1}{20}\mathcal{K}_3\bigg) \\
\mathcal{M}_9&=\delta_{\sigma}\bigg(-\frac{1}{30}\mathcal{K}_1\bigg),\mathcal{M}_{10}=\delta_{\sigma}\bigg(\frac{1}{300}\mathcal{K}_1-\frac{1}{20}\mathcal{K}_9\bigg)\nonumber,
\end{align}
  $\sigma$ denotes infinitesimal Weyl transformation parameter, and $K_i$ curvature invariantes from (\ref{ks}).
\bibliography{bibliothek}

\end{document}